\documentclass[12pt,onecolumn]{revtex4}
\usepackage{amssymb,epsf}
\begin{document}

\title{Quasilocal Thermodynamics of Kerr de Sitter Spacetimes\\
and the dS/CFT Correspondence}
\author{M. H. Dehghani}
\address{Physics Department and Biruni  Observatory,
         Shiraz University, Shiraz 71454, Iran}
\begin{abstract}
We consider the quasilocal thermodynamics of rotating black holes
in asymptotic de Sitter spacetimes. Using the minimal number of
intrinsic boundary counterterms, we carry out an analysis of the
quasilocal thermodynamics of Kerr-de Sitter black holes for
virtually all possible values of the mass, rotation parameter and
cosmological constant that leave the quasilocal boundary inside
the cosmological event horizon. Specifically, we compute the
quasilocal energy, the conserved charges, the temperature and the
heat capacity for the $(3+1)$-dimensional Kerr-dS black holes. We
perform a quasilocal stability analysis and find phase behavior
that is commensurate with previous analysis carried out through
the use of Arnowitt-Deser-Misner (ADM) parameters. Finally, we
investigate the non-rotating case analytically.
\end{abstract}

\maketitle


\section{Introduction}

\label{Intro}

The recent astrophysical data which indicate a positive
cosmological constant \cite{Perl} and the recent dS conformal
field theory (CFT) correspondence \cite{Str1,Odin} are two of the
main motivations for studying the thermodynamics of asymptotic de
Sitter (dS) black holes \cite{Cai}. Since these black holes have a
cosmological event horizon and there is no well-known
thermodynamic for outside the cosmological horizon, it is
worthwhile to study the quasilocal thermodynamic of asymptotic de
Sitter black holes inside the cosmological event horizon.

Among the quantities associated with gravitational thermodynamics
are the physical entropy $S$ and the temperature $\beta ^{-1}$,
where these quantities are, respectively, proportional to the area
and surface gravity of the event horizon(s) \cite{Beck,Haw1}. One
of the surprising and impressive features of this area law of
entropy is its universality. It applies to all kinds of black
holes and black strings \cite{HH,HHP,M1,M2}. It also applies to
the cosmological event horizon of the asymptotic de Sitter black
holes \cite{GH1}. Other black hole properties such as energy,
angular momentum and conserved charge, can also be given a
thermodynamic interpretation \cite{GH2}. But as is known, these
quantities typically diverge for asymptotic flat, AdS, and dS
spacetimes.

A common approach to evaluating them has been to carry out all
computations relative to some other spacetime that is regarded as
the ground state for the class of spacetimes of interest. This
could be done by taking the original action for gravity coupled to
matter fields and subtracting from it a reference action, which is
a functional of the induced metric $\gamma $ on the boundary
$\partial \mathcal{M}$. Conserved and/or thermodynamic quantities
are then computed relative to this boundary, which can then be
taken to infinity if desired.

This approach has been widely successful in providing a
description of gravitational thermodynamics in regions of both
finite and infinite spatial extent \cite{BY,BCM}. Unfortunately,
it suffers from several drawbacks. The choice of reference
spacetime is not always unique \cite{CCM}, nor is it always
possible to embed a boundary with a given induced metric into the
reference background. Indeed, for Kerr spacetimes this latter
problem creates a serious barrier against calculating the
subtraction energy, and calculations have only been performed in
the slow-rotating regime \cite{Mart}.

An extension of this approach which addresses these difficulties
was developed based on the conjectured AdS/CFT correspondence for
asymptotic AdS spacetimes \cite{BK,EJM} and recently for
asymptotic de Sitter spacetimes \cite {Bal1,GM1,Deh1}. It is
believed that appending a counterterm, $I_{ct}$, to the action
which depends only on the intrinsic geometry of the boundary(ies)
can remove the divergences. This requirement, along with general
covariance, implies that these terms are functionals of curvature
invariants of the induced metric and have no dependence on the
extrinsic curvature of the boundary(ies). An algorithmic procedure
exists for constructing $I_{ct}$\ \ for asymptotic AdS \cite{kls}
and dS spacetimes \cite{GM1}, and so its determination is unique.
Addition of $I_{ct}$\ will not affect the bulk equations of
motion, thereby eliminating the need to embed the given geometry
in a reference spacetime. Hence thermodynamic and conserved
quantities can now be calculated intrinsic for any given
asymptotically AdS or dS spacetime. The efficiency of this
approach has been demonstrated in a broad range of examples for
both the asymptotic AdS and dS spacetimes \cite{DaM,GM1,Deh1}.

Recently, we have considered the effects of including $I_{ct}$\
for quasilocal gravitational thermodynamics of Kerr and Kerr-AdS
black holes situations in which the region enclosed by $\partial
\mathcal{M}$ was spatially finite and we performed a quasilocal
stability analysis and found phase behavior that was commensurate
with a previous analysis carried out at infinity \cite{Deh2}. In
this paper, we investigate the thermodynamic properties of the
class of Kerr-dS black holes in the context of a spatially finite
boundary with radius less than the radius of the cosmological
horizon of the black hole, with the boundary action supplemented
by $I_{ct}$. With their lower degree of symmetry relative to
spherically symmetric black holes, these spacetimes allow for a
more detailed study of the consequences of including $I_{ct}$ for
spatially finite boundaries. There are several reasons for
considering this. First, there is no good interpretation of the
thermodynamics of the hole outside the cosmological horizon and
thereby at spatial infinity. Furthermore, the inclusion of
$I_{ct}$ eliminates the embedding problem from consideration,
whether or not the spatially infinite limit is taken, and so it is
of interest to see what its impact is on quasilocal
thermodynamics.

The outline of our paper is as follows. We review the basic
formalism in Sec. \ref{Genfor}. In Sec. \ref{Kerr} we consider the
Kerr-dS$_4$ spacetime and compute the entropy associated with the
cosmological event horizon by generalizing the Gibbs-Duhem
relation to this situation. In the next sections, we compute the
quasilocal energy, mass, angular momentum, temperature, and heat
capacity for Kerr-de Sitter black holes. These quantities must be
computed numerically for general values of the parameters, and we
present our results graphically. We perform a simple stability
analysis and compare to the previous analysis using the
Arnowitt-Deser-Misner (ADM) parameters \cite{Davies}. The
nonrotating case is considered in Sec. \ref{Sch}. We finish the
paper with some concluding remarks.

\section{General Formalism\label{Genfor}}

The gravitational action of asymptotic de Sitter spacetimes in
four dimensions is
\begin{equation}
I_G=-\frac 1{16\pi }\int_{\mathcal{M}}d^4x\sqrt{-g}\left( \mathcal{R}%
-2\Lambda \right) +\frac 1{8\pi }\int_{\partial \mathcal{M}}d^3x\sqrt{%
-\gamma }\Theta (\gamma ),  \label{IG}
\end{equation}
\ where the first term is the Einstein-Hilbert volume (or bulk) term with
positive cosmological constant $\Lambda =3/l^2$, and the second term is the
Gibbons-Hawking boundary term which is chosen such that the variational
principle is well-defined. The Euclidean manifold $\mathcal{M}$ has metric $%
g_{\mu \nu }$, covariant derivative $\nabla _\mu $, and time coordinate $%
\tau $ which foliates $\mathcal{M}$ into nonsingular hypersurfaces $\Sigma
_\tau $ with unit normal $u_\mu $ over a real line interval $\Upsilon $. $%
\Theta $ is the trace of the extrinsic curvature $\Theta ^{\mu \nu
}$ of any boundary(ies) $\partial \mathcal{M}$ of the manifold
$\mathcal{M}$, with induced metric(s) $\gamma _{ij}$. In general,
$I_G$ of Eq. (\ref{IG}) is divergent when evaluated on solutions,
as is the Hamiltonian and other associated conserved quantities.
Rather than eliminating these divergences by incorporating a
reference term in the spacetime, a new term $I_{ct}$ is added to
the action which is a functional only of boundary curvature
invariants \cite{Deh1}. Although there may exist a very large
number of possible invariants, one could add only a finite number
of them in a given dimension.\ Quantities such as energy, mass,
etc. can then be understood as intrinsically defined for a given
spacetime, as opposed to being defined relative to some abstract
(and nonunique) background, although this latter option is still
available. In four dimensions the counterterm is \cite{Deh1}
\begin{equation}
I_{ct}=\frac 2l\frac 1{8\pi }\int_{\partial \mathcal{M}_\infty
}d^3x \sqrt{-\gamma }\left( 1-\frac{l^2}4\mathcal{R}(\gamma
)\right) ,  \label{Ict}
\end{equation}
where $\mathcal{R}$ is the Ricci scalar of the boundary metric
$\gamma _{ab}$. We shall study the implications of including the
terms in (\ref{Ict}) for Kerr-dS black holes in spatially finite
regions. Although other counterterms (of higher mass dimension)
may be added to $I_{ct}$, they will make no contribution to the
evaluation of the action or Hamiltonian due to the rate at which
they decrease toward infinity, and we shall not consider them in
our analysis here.

Thorough discussion of the quasilocal formalism has been given
elsewhere \cite{BY,BCM,BM,Bal1} and so we only briefly
recapitulate it here. The action is the linear combination of the
gravity action $I_G$ (\ref{IG}) and the counterterm\ $I_{ct}$
given by (\ref{Ict}). Under the variation of the metric, one
obtains

\begin{equation}
\delta I=[\hbox{terms that vanish when the equations of motion hold}]^{\mu
\nu }\delta g_{\mu \nu }+\int_{\partial \mathcal{M}}d^3x(P^{ij}+Q^{ij})%
\delta \gamma _{ij},  \label{varact}
\end{equation}
where $P^{ij}$ is related to the variation of the Hawking-Gibbons boundary
term,
\begin{equation}
P^{ij}=\frac{\sqrt{-\gamma }}{8\pi }(\Theta ^{ij}-\Theta \gamma ^{ij}),
\label{stres1}
\end{equation}
\ and $Q^{ij}$ is due to the variation of the counterterm (\ref{Ict}) given
as

\begin{equation}
Q^{ij}=\frac{\sqrt{-\gamma }}{8\pi }\left\{-\frac{n-1}l\gamma
^{ij}+\frac l{n-2}(R^{ij}-\frac 12R\gamma ^{ij})\right\}.
\label{stres2}
\end{equation}

Decomposing the metric $\gamma _{ij}$ on the timelike boundary (the radius
of the boundary is less than the cosmological horizon and therefore the
boundary is timelike) $\partial \Sigma _\tau \times \Upsilon =\mathcal{B}%
\times \Upsilon =\mathcal{T}$ which connects the initial and final
hypersurfaces
\begin{equation}
\gamma _{ij}dx^idx^j=-N^2dt^2+\sigma _{ab}\left( dx^a+V^a\right) \left(
dx^b+V^b\right) ,  \label{metr2}
\end{equation}
yields after some manipulation
\begin{equation}
\left . \delta I\right| _{\mathcal{T}}=\int_{\mathcal{T}}d^3x\sqrt{\sigma }%
(-\varepsilon \delta N+j_a\delta V^a+\frac 12Ns^{ab}\delta \sigma
_{ab}),  \label{varact2}
\end{equation}
where the coefficients of the varied fields are
\begin{eqnarray}
\varepsilon  &=&\frac 2{N\sqrt{\sigma }}(P^{ij}+Q^{ij})u_iu_j,
\label{energy} \\
j_a &=&-\frac 2{N\sqrt{\sigma }}\sigma _{ai}(P^{ij}+Q^{ij})u_j,  \label{angm}
\\
\ s^{ab} &=&\frac 2{N\sqrt{\sigma }}\sigma _{\;i}^a\sigma
_{\;j}^b(P^{ij}+Q^{ij}),  \label{stres}
\end{eqnarray}
and we see that the counterterm variation $Q^{ij}$ supplants terms which
come from a reference action in the original formulation of the quasilocal
technique. Since $-\sqrt{\sigma }\varepsilon $ is the time rate of change of
the action, $\varepsilon $ is identified with the energy density on the
surface $\mathcal{B}$, and the total quasilocal energy for the system is
therefore
\begin{equation}
E=\int_{\mathcal{B}}d^2x\sqrt{\sigma }\varepsilon ,  \label{entot}
\end{equation}
and this quantity can be meaningfully associated with the thermodynamic
energy of the system \cite{CrThes}. Using similar reasoning, the quantities $%
j_a$ and $s^{ab}$ are, respectively, referred to as the momentum surface
density and the spatial stress.

When there is a Killing vector field $\mathcal{\xi }$\ on the boundary $%
\mathcal{T}$ , an associated conserved charge is defined by
\begin{equation}
\mathcal{Q}\left( \mathcal{\xi }\right) =\int_{\mathcal{B}}d^2x\sqrt{\sigma }%
\left( \varepsilon u^i+j^i\right) \mathcal{\xi }_i,  \label{cons}
\end{equation}
provided there is no matter stress energy in the neighborhood of
$\mathcal{T} $ (this assumption can be dropped, allowing one to
compute the time rate of change of $\mathcal{Q}$ if desired
\cite{CrBo}). When this holds, the
value of $\mathcal{Q}$ is independent of the particular hypersurface $%
\mathcal{B}$ , a property not shared by the energy $E$. For
boundaries with timelike [$\xi ^a=(\partial /\partial t)^a/\Xi $]
and rotational Killing vector fields ($\varsigma =\partial
/\partial \phi $), we obtain
\begin{eqnarray}
M &=&\int_{\mathcal{B}}d^2x\sqrt{\sigma }\left( \varepsilon u^i+j^i\right)
\xi _i,  \label{Mastot} \\
J &=&\int_{\mathcal{B}}d^2x\sqrt{\sigma }j^i\varsigma _i,  \label{Angtot}
\end{eqnarray}
provided the surface $\mathcal{B}$ contains the orbits of
$\varsigma $. These quantities are, respectively, the conserved
mass and angular momentum of the system enclosed by the boundary.
Note that they will both be dependent on the location of the
boundary $\mathcal{B}$ in the spacetime, although each is
independent of the particular choice of foliation $\mathcal{ B}$
within the surface $\mathcal{T}$.

In the context of the dS/CFT correspondence, the limit in which
the boundary $\mathcal{B}$ becomes infinite is taken, and the
counterterm prescription \cite{Deh1} ensures that the conserved
charges (\ref{cons}) are finite. No embedding of the surface
$\mathcal{T}$ \ into a reference spacetime is required. This is of
particular advantage for the class of Kerr spacetimes in which it
is not possible to embed an arbitrary two-dimensional boundary
surface into a flat (or constant-curvature) spacetime
\cite{Mart,Deh2}.

\section{The Kerr-dS$_4$ spacetime\label{Kerr}}

The class of metrics we shall consider are the Kerr-dS family of solutions
in four dimension, whose general form is
\begin{eqnarray}
ds^2 &=&-\frac{\Delta _r^2}{\rho ^2}\left(dt-\frac a\Xi \sin
^2\theta d\phi \right)^2+ \frac{\rho ^2}{\Delta _r}dr^2+\frac{\rho
^2}{\Delta _\theta }d\theta ^2
 \nonumber\\
&&+\frac{\Delta _\theta ^2\sin ^2\theta }{\rho
^2}\left(adt-\frac{(r^2+a^2)}\Xi d\phi \right)^2, \label{met1a}
\end{eqnarray}
where
\begin{eqnarray}
\Delta _r^2 &=&(r^2+a^2)(1-r^2/l^2)-2mr,  \nonumber \\
\Delta _\theta ^2 &=&1+\frac{a^2}{l^2}\cos ^2\theta ,  \nonumber\\
\Xi  &=&1+\frac{a^2}{l^2},  \nonumber \\
\rho ^2 &=&r^2+a^2\cos ^2\theta . \label{met1b}
\end{eqnarray}
For $m=0$, the metric (\ref{met1a} and \ref{met1b}) is that of
pure dS$_4$ spacetime (or flat spacetime if $l\rightarrow \infty
$), and for $a=0$ the metric is that of Schwarzschild-dS$_4$
spacetime, which has zero angular momentum. \ Hence we expect the
parameters $m$ and $a$ to be associated with the mass and angular
momentum of the spacetime, respectively. The metric of Eqs.
(\ref{met1a}) and (\ref{met1b}) has three horizons located at
$r_{-}$, $ r_{+}$, and $r_c$, provided the parameters $m$, $l$,
$a$ are chosen suitably. For given values of $l$ and $a$, $m$
should lie in the range between $ m_{1,crit}\leq m\leq
m_{2,crit}$, where $m_{1,crit}$ and $m_{2,crit}$ are the positive
real solutions of the following equation:

\begin{equation}
a^{10}+4l^2a^8+4l^6a^4+6l^4a^6-33a^2l^6m^2+33a^4l^4m^2+27m^4l^6-m^2l^8+m^2l^2a^6=0.
\label{Crit}
\end{equation}
It is worthwhile to mention that when $m=m_{1,crit}$, $r_{-}$ and
$r_{+}$
are equal and we have an extreme black hole. For the case in which $%
m=m_{2,crit}$, the radius of all three horizons are equal and we
have a critical hole. Also, one may note that in the limit
$l\rightarrow \infty $, $m_{1,crit}=a$, $m_{2,crit}\rightarrow
\infty $, $r_{\pm }=m\pm \sqrt{m^2-a^2} $, and $r_c\rightarrow
\infty $. For given values of $a$ and $m$, the parameter $l$
should be greater than $l_{crit}$, where $l_{crit}$ is the
positive real solution of Eq. (\ref{Crit}). For given values of
$l$ and $m$, the parameter $a$ should be less than a critical
value $a_{crit}$, where it is the largest real value solution of
Eq. (\ref{Crit}).

The total action of the system, $I=I_G+I_{ct}$, can be calculated
through the use of Eqs. (\ref{IG}) and (\ref{Ict}):
\begin{equation}
I=-\frac{\beta _c}{2\Xi l^2}[ml^2+r_c(r_c^2+a^2)],  \label{Itot}
\end{equation}
where $r_c$ is the radius and $\beta _c$ is the inverse of the
Hawking temperature of the cosmological event horizon given by
\begin{equation}
\beta _c=-\frac{4\pi l^2r_c(r_c^2+a^2)}{3r_c^4+a^2r_c^2-l^2r_c^2+a^2l^2}.
\label{betc}
\end{equation}
The conserved mass and angular momentum of the hole calculated on the
boundary $\mathcal{B}$ at infinity calculated from Eqs. (\ref{cons}) are
given by \cite{Deh1}:
\begin{eqnarray}
M_\infty  &=&-\frac m\Xi ,  \label{Mto} \\
J &=&\frac{ma}{\Xi ^2}.  \label{Jto}
\end{eqnarray}
Now generalizing the Gibbs-Duhem relation for asymptotic de Sitter black
holes \cite{HHP},

\begin{equation}
S=I-\beta _c(M+\Omega J),  \label{ent}
\end{equation}
the entropy associated with the cosmological event horizon can be calculated
as

\begin{equation}
S=\frac{\pi (r_c^2+a^2)}\Xi .  \label{Sc}
\end{equation}
As we see, the entropy (\ref{Sc}) is in agreement with the area
law of entropy even for the cosmological horizon \cite{GH1}.

\section{The quasilocal thermodynamics of the black hole\label{Ther}}

Now we consider the quasilocal thermodynamics of Kerr-dS$_4$ black
holes. The energy of the system can be written as
\begin{equation}
E=E_1+E_2;\ \ \ \ \ \ \ \ \ \ \ \ \ \ \ \ \ \ E_i=\int_{\mathcal{B}}d^2x%
\sqrt{\sigma }\varepsilon _i,  \label{Et}
\end{equation}
where $\mathcal{B}$ is a two dimensional surface defined by setting the
radial coordinate to a constant value $r<r_c$, and $\varepsilon _1$ and $%
\varepsilon _2$ are given by\
\begin{equation}
\sqrt{\sigma }\varepsilon _1=\frac 2NP^{ij}u_iu_j=\frac{r\Delta _r\left\{\left(1+\frac{%
a^2}{l^2}-\frac mr\right)a^2\cos ^2\theta +\left(1+\frac{a^2}{l^2}+\frac mr+2\frac{r^2}{%
l^2}\right)a^2+2r^2\right\}}{8\pi \rho \Xi \Delta _\theta
[(r^2+a^2)^2\Delta _\theta ^2-a^2\Delta _r^2\sin ^2\theta
]^{1/2}}\sin \theta ,  \label{Eden1}
\end{equation}

\begin{eqnarray}
\sqrt{\sigma }\varepsilon _2 &=&\frac 2NQ^{ij}u_iu_j=\frac{\sin \theta }{%
8\pi l\rho ^6\Xi \Delta _\theta [(r^2+a^2)^2\Delta _\theta ^2-a^2\Delta
_r^2\sin ^2\theta ]^{1/2}}\times   \nonumber \\
&&\ \{[\Xi (r^2+a^2)-2mr]a^8\cos ^8\theta +[2\Xi
r^2(r^2+a^2)-mr(3l^2-a^2)]a^6\cos ^6\theta   \nonumber \\
&&\ +[\Xi r^2(r^2+a^2)(7r^2-l^2)+2mr(r^4+2ml^2r-l^2a^2+11r^2a^2)]a^4\cos
^4\theta   \nonumber \\
&&\ +[2\Xi r^4(r^2+a^2)(3r^2-l^2)+2mr(4l^2a^4-2r^2a^4+7r^2l^2a^2+5l^2r^4
\nonumber \\
&&\ +2a^2r^4-2mra^2l^2-2r^6)]a^2\cos ^2\theta +\Xi r^6(r^2+a^2)(2r^2-l^2)-
\nonumber \\
&&\ 2ma^2r^3(l^2a^2+l^2r^2-2r^4)\}.  \label{Eden2}
\end{eqnarray}

We first consider special cases in which the integral in Eqs.
(\ref{Eden1}) and (\ref{Eden2}) can be evaluated. For $r=r_{+}$,
where $r_{+}$ is the radius of the outer event horizon, we find
that the total energy can be explicitly computed, yielding
\begin{equation}
E(r_{+})=\frac l{2\Xi
}\left(1-\frac{a^2}{l^2}-2\frac{r_{+}^2}{l^2}\right). \label{Ehor}
\end{equation}
For small values of angular momentum, $E$ can be integrated
easily. Then the total energy for small $a$ is
\begin{eqnarray}
E &=&\frac
l2\left(1-2\frac{r^2}{l^2}\right)-r\sqrt{1-\frac{r^2}{l^2}-2\frac
mr}
- \{\frac 2{3l}\left(1-\frac{r^2}{l^2}+\frac mr\right)  \nonumber \\
&&+\frac 1{3r}\left (1-\frac{r^2}{l^2}-2\frac mr \right )^{-1/2}
\left(1-3\frac{r^2}{l^2}+2\frac{r^4}{l^4}+5\frac{mr}{l^2}+2\frac{m^2}{r^2}\right)
\}a^2+O(a^4). \label{Esa}
\end{eqnarray}

\begin{figure}
  \epsfxsize=10cm
  \centerline{\epsffile{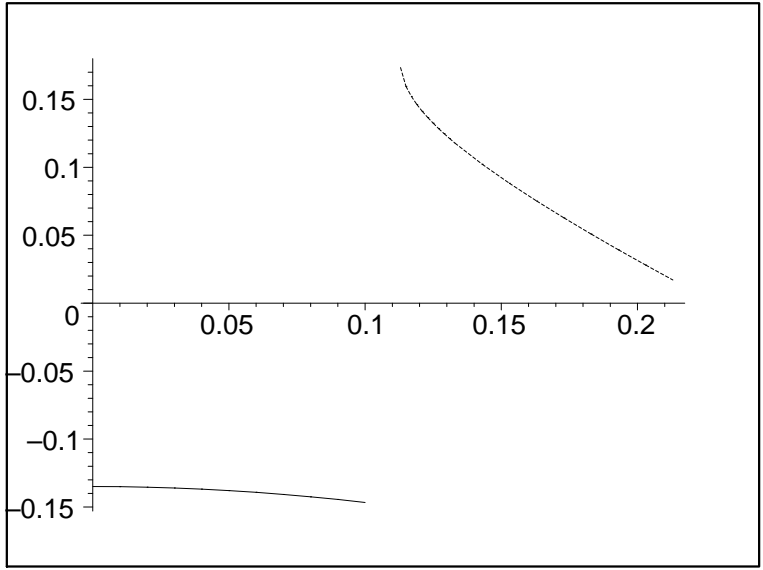}}
  \caption{ $E$
versus $a$ for $l=1$, $m=0.1$ (solid), and $0.2$ (dashed).}
  \label{Figure1}
\end{figure}

\begin{figure}\
  \epsfxsize=10cm
  \centerline{\epsffile{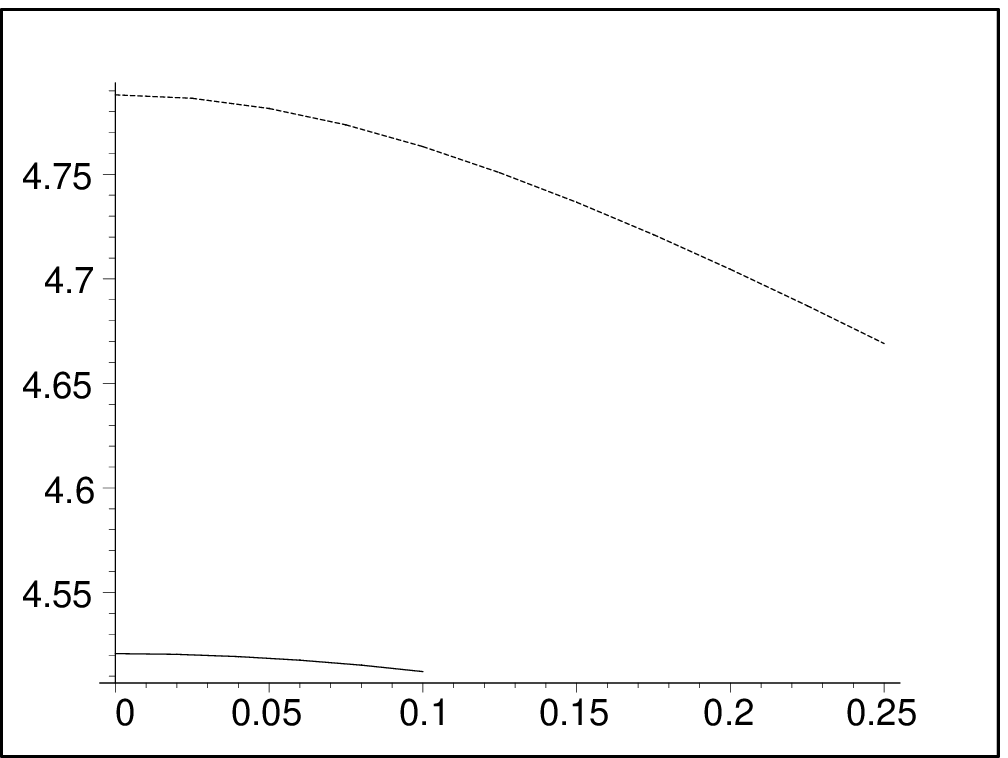}}
  \caption{ $E$
versus $a$ for $l=10$, $m=0.1$ (solid), and $0.2$ (dashed).}
  \label{Figure2}
\end{figure}

\begin{figure}
  \epsfxsize=10cm
  \centerline{\epsffile{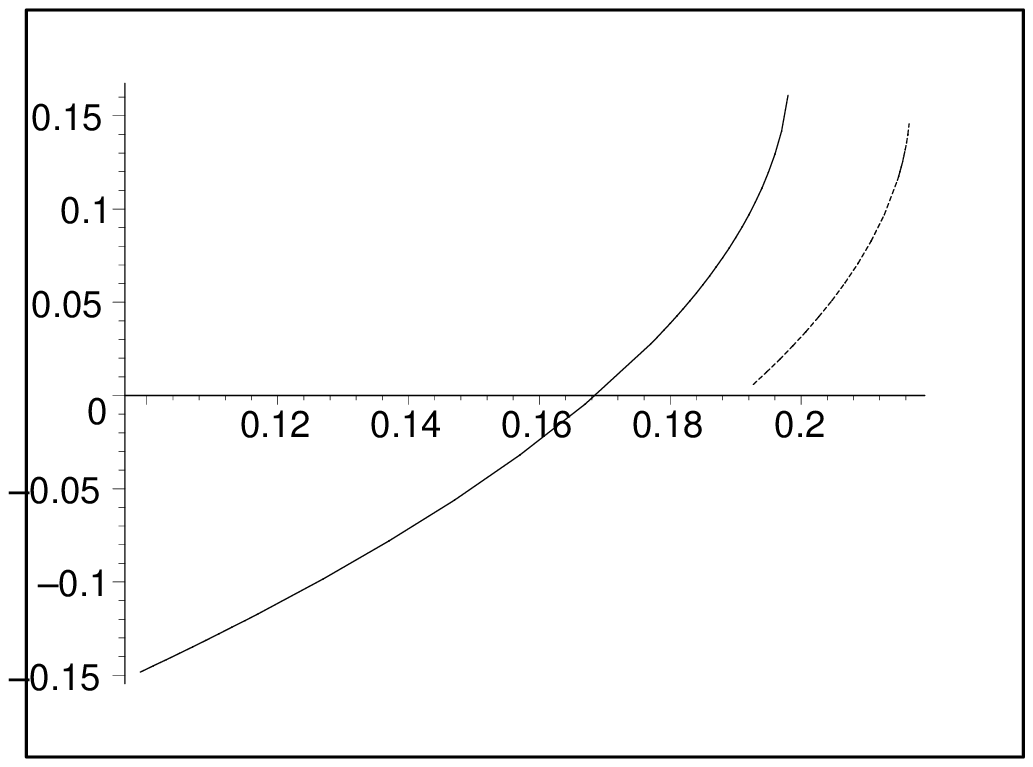}}
  \caption{ $E$
versus $m$ for $l=1$, $a=0.1$ (solid), and $0.2$ (dashed).}
  \label{Figure3}
\end{figure}

\begin{figure}
  \epsfxsize=10cm
  \centerline{\epsffile{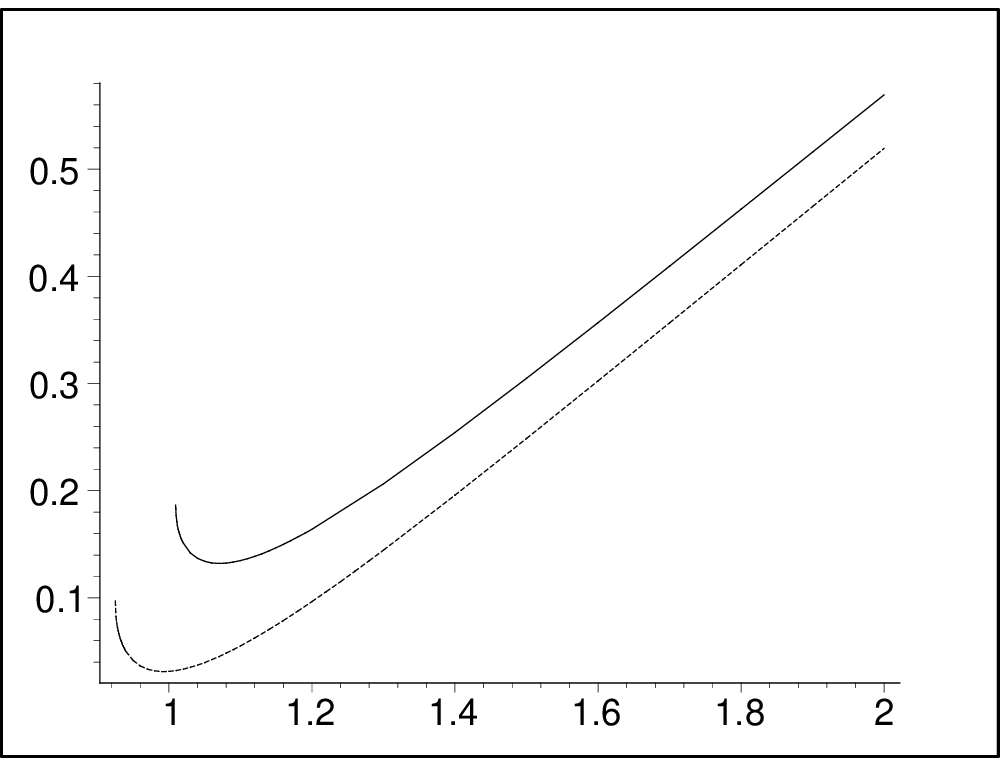}}
  \caption{ $E$
versus $l$ for $m=0.2$, $a=0.1$ (solid), and $0.2$ (dashed).}
  \label{Figure4}
\end{figure}

For general values of the parameters, however, we cannot
analytically integrate (\ref{Eden1}) and (\ref{Eden2}) exactly,
and so we resort to numerical integration. To do this we set the
radius of the boundary equal to $r=0.56$, and we apply the
restrictions discussed in Sec. \ref{Kerr} on the parameters $a$,
$m$ and $l$. Note that for $r=0.56$, each of these parameters has
two critical values, one due to the extreme Kerr-dS black hole and
the other due to the case in which this radius is equal to the
radius of the three event horizons (in this case the radius of the
three event horizons are equal). In the first four figures we plot
the quasilocal energy as a function of $a$%
, $m$, and $l$. The $a$ dependence of the energy is given in Figs.
\ref {Figure1} and \ref{Figure2}. We see that for small values of
\ $l$, the
energy is a slowly increasing function of $a$, while for larger values of $l$%
, the energy is a decreasing function of $a$ as in the case of the
Kerr metric
considered in \cite{Deh2}. Figures \ref{Figure3} and \ref{Figure4} show the $%
m$ and $l$ dependence of the energy. For large values of $l$, the
energy is asymptotic to a linear function of $l$. This is due to
the fact that the counterterm used for the Kerr-dS spacetime is
proportional to $l$ for large values of this parameter.

Next we consider the quasilocal conserved charges of the Kerr-dS
metric, which are the mass and $\varphi $ component of the angular
momentum of the system defined in Eqs. (\ref{Mastot}) and
(\ref{Angtot}). The $\varphi $ component of the angular momentum
due to the counterterm is zero and one can evaluate the total
angular momentum as

\begin{equation}
J=\frac{am}{\Xi ^2} ,  \label{Jt}
\end{equation}
which is therefore valid for any quasilocal surface of fixed radius $r$, and
yields the angular momentum of the Kerr metric as $l$ goes to infinity.

Using the normalized Killing vector $\zeta ^a=(\partial /\partial t)^a/$ $%
\Xi $ instead of $(\partial /\partial t)^a$ in Eq. (\ref{Mastot}), the total
mass of the system can be integrated:

\begin{eqnarray}
M &=&\frac 1{\Xi ^2}\{\frac{\Delta _r^2}{2a}\tan ^{-1}\left(\frac
ar\right)+\frac r2\left(1-\frac{r^2}{l^2}-2\frac mr\right)
-\frac{r^2\Delta _r}{8la}\ln \left(\frac{\sqrt{r^2+a^2}%
+a}{\sqrt{r^2+a^2}-a}\right)  \nonumber  \\
&&-\Delta _r\left[\frac{a^2+3r^2-2l^2}{4l(r^2+a^2)^{1/2}}+\frac{mla^2}{%
3r(r^2+a^2)^{3/2}}\right]\}.  \label{Mt}
\end{eqnarray}
For $r=r_{+}$,\ one obtains

\begin{equation}
M(r_{+}) =\frac{r_{+}}{2\Xi ^2}\left(1-\frac{r_{+}^2}{l^2}-2\frac
m{r_{+}}\right)=\frac{a^2}{2\Xi
^2r_{+}}\left(1-\frac{r_{+}^2}{l^2}\right). \label{Mhor}
\end{equation}
Note that in the limit of slow rotation, $M\left( r_{+}\right) $
vanishes, a strictly different feature from that of the energy as
given in Eq. (\ref {Ehor}), which in this limit equals
$(l/2)[1-2(r_{+}^2/l^2)]$.

We turn next to a thermodynamics consideration of the Kerr-dS
black hole. The interior of the quasilocal surface can be regarded
as a thermodynamic system, which is corotating with an exterior
heat bath whose angular velocity is $\Omega _{+}$, where $\Omega
_{+}=a(r_{+}^2+a^2)^{-1}$ is the angular velocity of the outer
event horizon. The quasilocal thermodynamic quantities are
generally given by surface integrals over the boundary data rather
than products of constant surface data. In particular, it is not
possible to choose the quasilocal surface to simultaneously be
both isothermal and a surface of constant $\Omega $, nor is this
possible for the quasilocal surface we have chosen (one at fixed
$r$). \ Furthermore, it is by now well-recognized that the entropy
of a black hole depends only on the geometry of its horizon in the
classical approximation, and is conversely independent of the
asymptotic behavior of the gravitational field or of the presence
of external matter fields. For stationary black holes, the entropy
is one-quarter of the horizon area, yielding\ $S=\pi (
r_{+}^2+a^2) /\Xi $ in this case. We shall therefore regard the
energy $E$ in Eq. (\ref{Et}) as the thermodynamic internal energy
for the Kerr black hole within the boundary with radius $r$, view
it as a function of $r$, the entropy $S$, and the angular momentum
$J$, and define the temperature and the angular velocity from
their ensemble derivatives, integrated over the quasilocal
boundary data.

For the temperature we obtain
\begin{equation}
T=\left( \frac{\partial E}{\partial S}\right) _{r,J}=\left( \frac{\partial E%
}{\partial M}\right) _{r,J}\left( \frac{\partial S}{\partial M}\right)
_{r,J}^{-1}=\beta _{+}\left( \frac{\partial E}{\partial M}\right) _{r,J},
\label{Tem}
\end{equation}
where $\beta _{+}$\ is the inverse of Hawking temperature
associated with the outer horizon,
\begin{equation}
\beta _{+}=-\frac{3r_{+}^4+r_{+}^2(a^2-l^2)+a^2l^2}{4\pi
l^2r_{+}(r_{+}^2+a^2)}.  \label{bet}
\end{equation}
It is worthwhile to mention that unlike the nonrotating case, the factor $%
\beta _{+}$ does not give the Tolman redshift factor because a
surface of fixed radius $r$ is not a surface of constant redshift
(it is not an isotherm). The derivative $(\partial E/\partial
M)_{r,J}$ is obtained by taking the derivatives of the terms in
Eqs. (\ref{Eden1}) and (\ref{Eden2}) above and then performing the
integration with respect to $\theta $.

\begin{figure}
  \epsfxsize=10cm
  \centerline{\epsffile{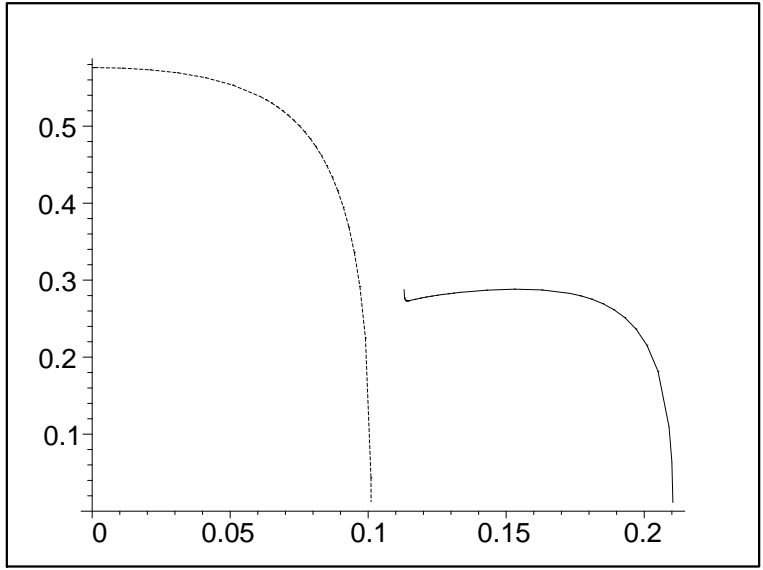}}
  \caption{ $T$
versus $a$ for $l=1$, $m=0.1$ (dashed), and $0.2$ (solid).}
  \label{Figure5}
\end{figure}

\begin{figure}
  \epsfxsize=10cm
  \centerline{\epsffile{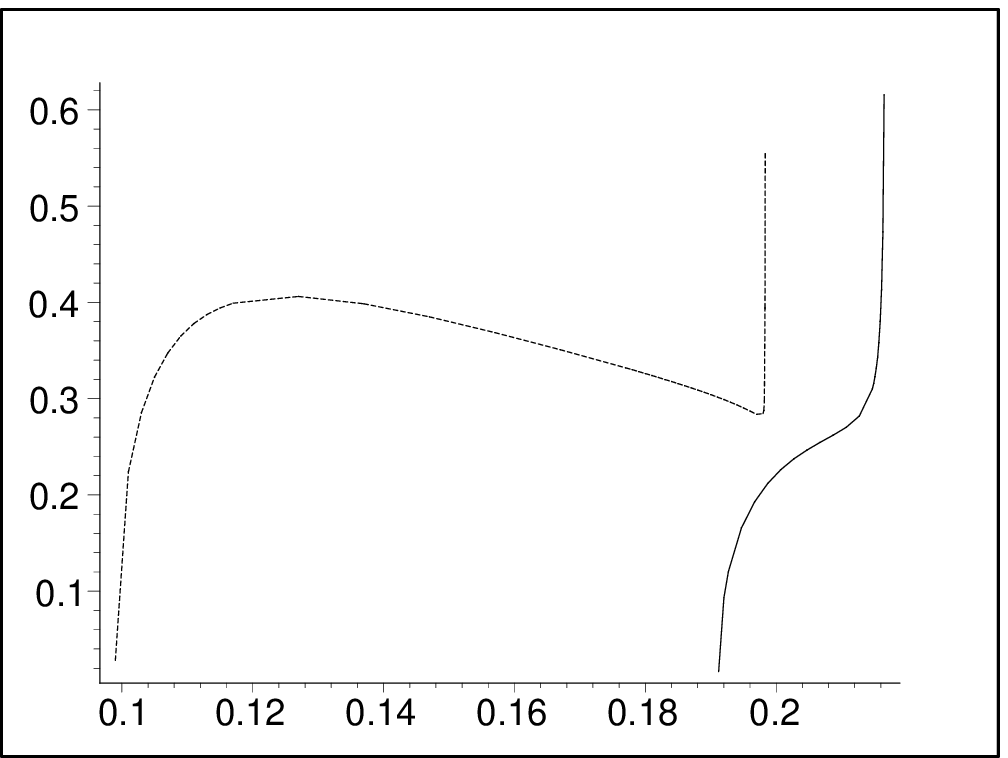}}
  \caption{ $T$
versus $m$ for $l=1$, $a=0.1$ (dashed), and $0.2$ (solid).}
  \label{Figure6}
\end{figure}

\begin{figure}
  \epsfxsize=10cm
  \centerline{\epsffile{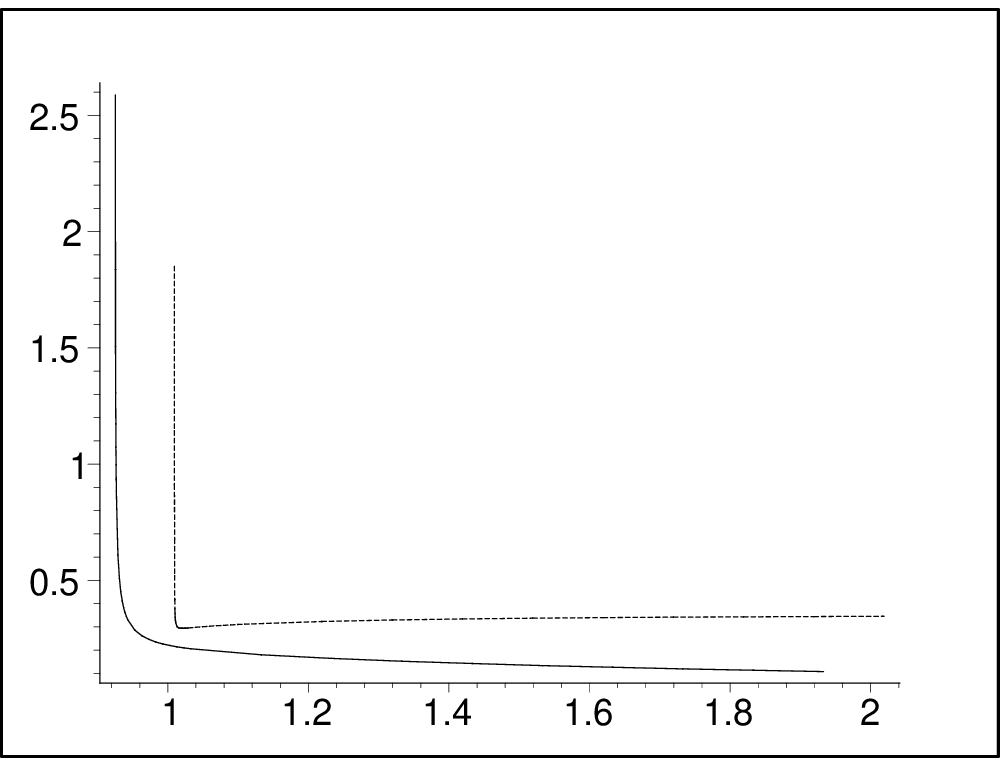}}
  \caption{ $T$
versus $l$ for $m=0.2$, $a=0.1$ (dashed), and  $0.2$ (solid).}
  \label{Figure7}
\end{figure}

\begin{figure}
  \epsfxsize=10cm
  \centerline{\epsffile{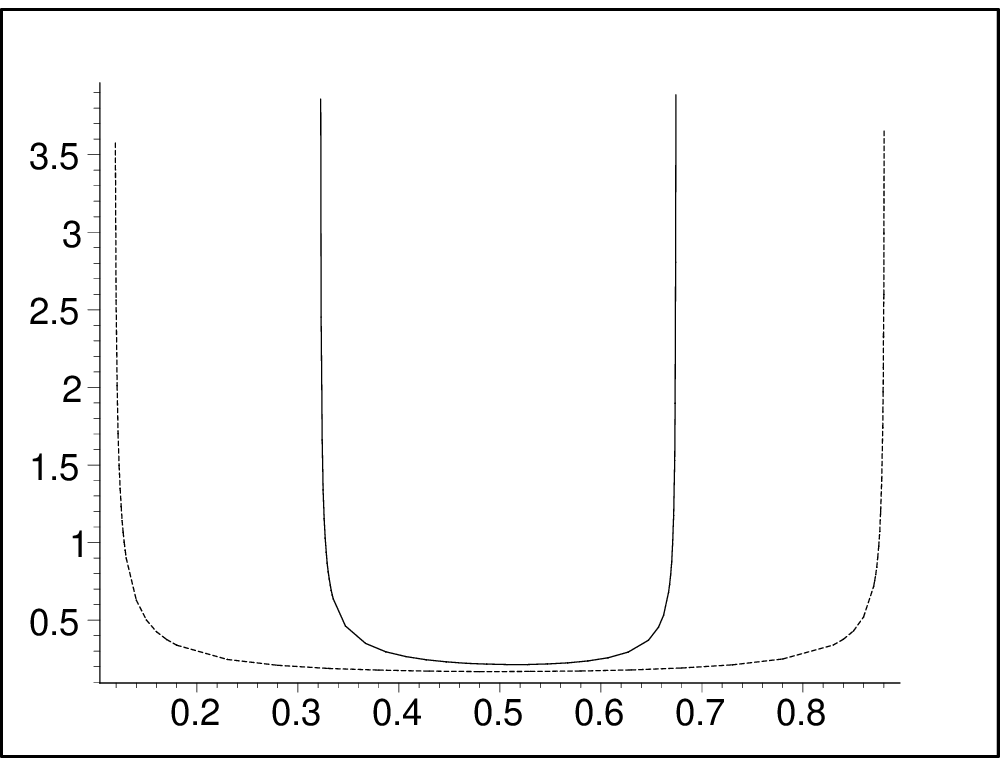}}
  \caption{ $T$
versus $r$ for $l=1$, $m=0.2$, $a=0.1$ (dashed), and $0.2$
(solid).}
  \label{Figure8}
\end{figure}

Now we investigate the $a$, $m$, and $l$ dependence of the
temperature of a Kerr-dS black hole. For the general case, the
temperature must be computed numerically. The $a$ dependence of
the temperature is shown in Fig. \ref {Figure5}. We find that for
small values of $m$, the temperature is a weakly decreasing
function of $a$, and it decreases rapidly as $a$ approaches its
maximum allowed value $a_{crit}$ (extreme black hole). For larger values of $%
m$ ($l=1$ and $m=0.2$), the temperature increases slowly as $a$
increases,
but again it goes to zero rapidly as $a$ approaches its critical value. The $%
m$ dependence of $T$\ is shown in Fig. \ref{Figure6}. We find that
for small values of $a$ ($a=0.1$) the function $T(m)$ has a
maximum at $m\simeq 0.116$ and a minimum at $m\simeq 0.188$ and
therefore $(\partial T/\partial M)_{r,J}$ is negative between
these two values. The $l$ and $r$ dependence of the temperature
can be seen from Figs. \ref{Figure7} and \ref{Figure8}. We find
that $T$ is an initially strongly decreasing function of $l$, but
as $l$ increases the temperature increases linearly. Note that for
$m=0.2$ and $a=0.2$ (a closely extreme hole), $T$ goes to zero as
$l$ goes to infinity, which is commensurable to our previous
results in \cite{Deh2}.

The heat capacity at constant surface area, $4\pi r^2$, is defined
by
\begin{equation}
C_r=\left( \frac{\partial E}{\partial T}\right) _{r,J},  \label{km16}
\end{equation}
where the energy is expressed as a function of $r$ and $T.$
Expressing the energy and temperature as functions of $r$, $M$,
and $J$, the heat capacity can be written as
\begin{equation}
C_r=\left( \frac{\partial E}{\partial M}\right) _{r,J}\left( \frac{\partial T%
}{\partial M}\right) _{r,J}^{-1}.  \label{Cap}
\end{equation}

\begin{figure}
  \epsfxsize=10cm
  \centerline{\epsffile{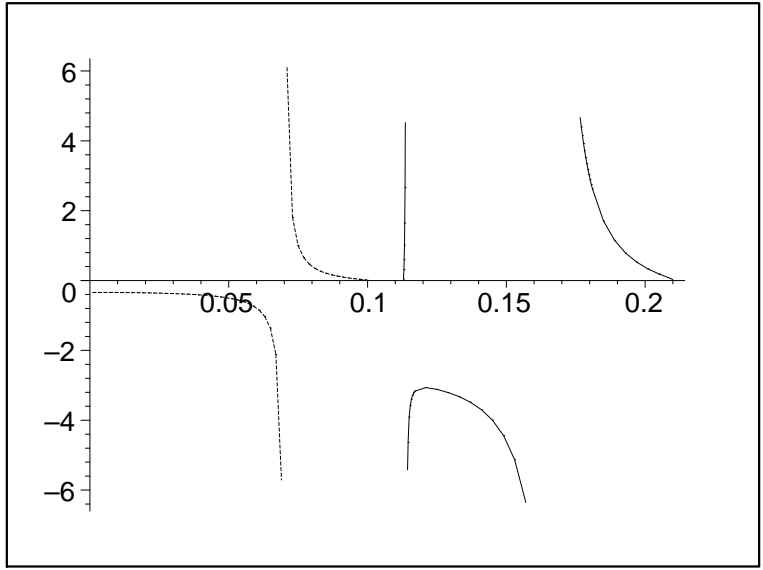}}
  \caption{ $C$
versus $a$ for $l=1$, $m=0.1$ (dashed), and $0.2$ (solid).}
  \label{Figure9}
\end{figure}

\begin{figure}
  \epsfxsize=10cm
  \centerline{\epsffile{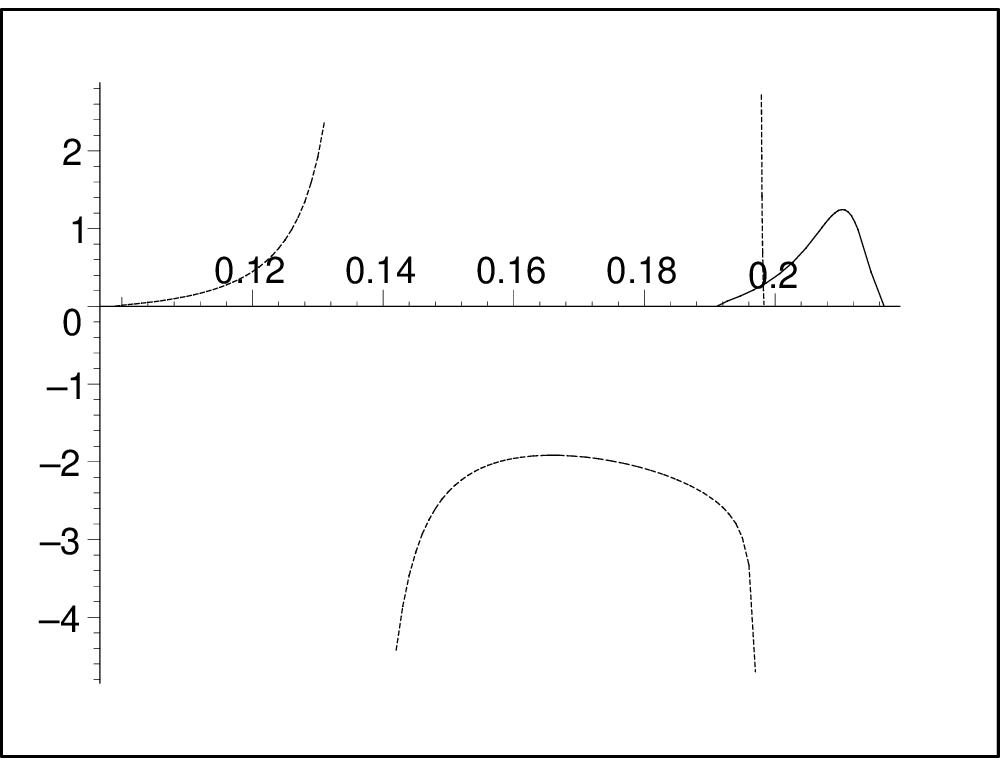}}
  \caption{ $C$
versus $m$ for $l=1$, $a=0.1$ (dashed), and $0.2$ (solid).}
  \label{Figure10}
\end{figure}

For small values of $m$, the heat capacity is not positive for all
the allowed values of $a$. For example, as one can see from Fig.
\ref{Figure9},
for $l=1$ and $m=0.1$ the heat capacity is positive only for $a\gtrsim 0.07$%
, while for larger values of $m$ ($m=0.2$) there are two stable phases, a
slowly rotating black hole ($a\lesssim 0.11$) and a nearly extreme hole ($%
0.17\lesssim m\lesssim 0.21$). These two stable phases are separated by an
intermediate angular momentum unstable phase ($0.11\lesssim m\lesssim 0.17$%
). Indeed, for this value of $m$ there exist two $a$'s for which
the heat capacity passes through an infinite discontinuity,
characteristic of a second-order phase transition. We see from
Fig. \ref{Figure10} that for small values of $a$ there is a phase
of stable small black holes and a phase of stable large black
holes, separated by a regime of intermediate unstable black holes.
However, for sufficiently large $a$, the heat capacity is positive
for all allowed values of $m$, and there is a single phase
consisting of a nearly extreme black hole. The stability analysis
for these black holes is in qualitative agreement with that of
Davies \cite{Davies}, who performed a thorough analysis of the
thermodynamic properties of asymptotic de Sitter Kerr-Newman black
holes. However, he used the ADM mass parameters, whereas we are
considering the quasilocal energy $E$ in Eq. (\ref{Et} ) as a
function of $r$, $S$, and $J$.

\begin{figure}
  \epsfxsize=10cm
  \centerline{\epsffile{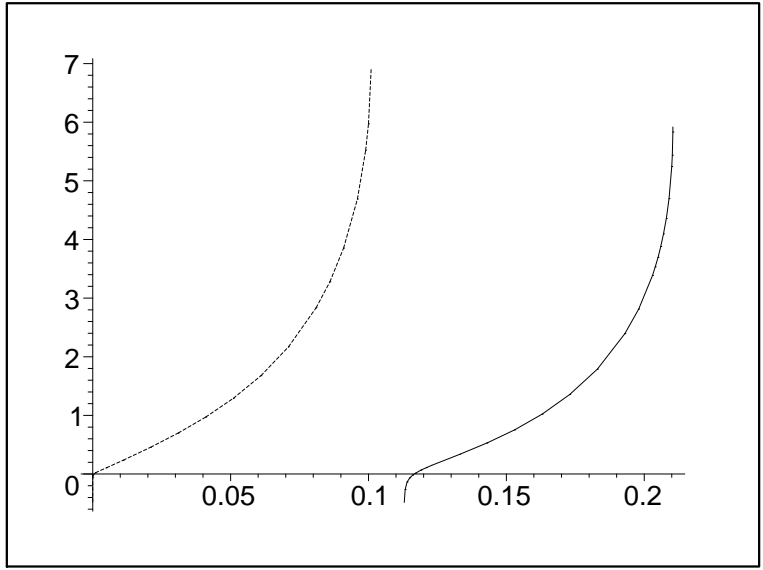}}
  \caption{ $\Omega$
versus $a$ for $l=1$, $m=0.1$ (dashed), and $0.2$ (solid).}
  \label{Figure11}
\end{figure}

The angular velocity can be written as
\begin{equation}
\Omega =\left( \frac{\partial E}{\partial J}\right) _{r,S}.  \label{Omeg}
\end{equation}
The $a$ dependence of the angular velocity can be seen from Fig.
\ref{Figure11}.

\section{The quasilocal thermodynamics of The Nonrotating Case\label{Sch}}

Now we investigate the nonrotating case in which all the
thermodynamic quantities can be integrated easily. For $a=0$, one
obtains the quasilocal energy of a Schwarzschild-dS black hole as

\begin{equation}
E=\frac
l2\left(1-2\frac{r^2}{l^2}\right)-r\sqrt{1-\frac{r^2}{l^2}-2\frac
mr}. \label{Esch}
\end{equation}

Using Eq. (\ref{Tem}), the temperature of the black hole is
\begin{equation}
T=\frac{1-3\frac{r_{+}^2}{l^2}}{4\pi r_{+}\sqrt{1-\frac{r^2}{l^2}-2\frac mr}}%
.  \label{Tsch}
\end{equation}
It is worthwhile to mention that the temperature given by Eq.
(\ref{Tsch}) can be obtained by a Wick rotation $l\rightarrow il$
from the expression given in Ref. \cite{BCM} for a Scwarzschild
AdS black hole. The heat capacity can be obtained from (\ref{Cap})
as
\begin{equation}
C_r=-\frac{4\pi r_{+}^2\left(1-\frac{r^2}{l^2}-2\frac mr\right)\left(1-3\frac{r_{+}^2}{l^2}\right)%
}{r_{+}\left(3-2\frac{r_{+}^2}{l^2}+3\frac{r_{+}^4}{l^4}\right)+2r\left(1+3\frac{r_{+}^2}{l^2%
}\right)\left(1-\frac{r^2}{l^2}\right)}.  \label{Csch}
\end{equation}
One may note that the heat capacity $C$ passes through an infinite
discontinuity, characteristic of a second-order phase transition,
when the denominator of the heat capacity vanishes, i.e.,
\begin{equation}
3r_{+}^5-2l^2r_{+}^3-6r(l^2-r^2)r_{+}^2+3l^4r_{+}-2rl^2(l^2-r^2)=0.
\label{phase}
\end{equation}
For a given $l$ and a fixed value of $r$ in the range
$r_{+}<r<r_c$, Eq. (\ref {phase}) has a single real solution for
$r_{+}$, which belongs to a value for the parameter $m$ denoted by
$m_0$. The heat capacity (\ref{Csch}) for black holes with $m>m_0$
is positive and therefore they are stable. Note that as
$l\rightarrow \infty $,  $m_0\rightarrow r/3$.

\section{Closing Remarks\label{Clos}}

In this paper, we have computed the entropy associated with the
cosmological event horizon of the Kerr-dS black hole by
generalizing the Gibbs-Duhem relation for asymptotic de Sitter
spacetimes. The result is in agreement with the well-known area
law of the entropy. Since there is no well-defined thermodynamics
for outside the cosmological horizon, an investigation of the
quasilocal thermodynamics is necessary. Our investigation of the
boundary-induced counterterm prescription (\ref{Ict}) at finite
distances has allowed us to obtain a number of interesting
results, which we recapitulate here.

We have been able to compute the energy, mass, angular momentum,
temperature, and heat capacity quasilocally for arbitrary values
of the parameters of the Kerr-dS$_4$ solution, apart from the mild
reality restrictions considered in Sec. \ref{Kerr}. These
quasilocal quantities are intrinsically calculable numerically at
fixed $r$ without any reference to a background spacetime, a
marked improvement over the methods given in Ref. \cite{BY}. We
also found that quasilocal angular momentum is independent of the
radius of the boundary, a fact which is not true for the total
mass of the spacetime. This remarkable result holds because the
angular momentum due to the counterterm integrates to zero, and
not because it vanishes identically.

We found that the entropy $S$ and angular momentum $J$ were not
given by surface integrals over quasilocal boundary data, but
rather were constants dependent on the parameters $a$, $m$ and $l$
of the black hole. We therefore chose to regard the interior of
the quasilocal surface as a thermodynamic system whose energy $E$
is a function of $S$ and $J$. We found that the temperature goes
to zero as the black hole approaches its extreme case, where the
radii of the inner and outer horizons are equal, and goes to
infinity as the radius of the boundary approaches the radius of
the outer or cosmological horizons.

A stability analysis yielded results that are in qualitative
agreement with previous investigations carried out by using the
ADM mass parameter \cite {Davies}. This stability analysis is, of
course, local, and phase transitions to other spacetimes of lower
free energy might exist. Finally, we have considered the case of
nonrotating black holes and found that at a fixed value of $r$
there exists a value for the mass parameter ($m_0$) for which the
heat capacity passes an infinite discontinuity. For $m>m_0$, the
black hole is stable while there exists a nonstable phase for
$m<m_0$.

\end{document}